\documentclass[11pt]{article}
\usepackage[russian]{babel}
\usepackage{amssymb,amsmath,amsthm}
\usepackage{graphicx}
\usepackage{epstopdf}
\newcommand{\bc}{\begin{center}}
\newcommand{\ec}{\end{center}}
\newcommand{\be}{\begin{equation}}
\newcommand{\ee}{\end{equation}}
\newcommand{\bea}{\begin{eqnarray}}
\newcommand{\eea}{\end{eqnarray}}
\newcommand{\ba}{\begin{array}}
\newcommand{\ea}{\end{array}}
\newcommand{\edc}{\end{document}}

\def\s{\sigma}

\begin{document}
\begin{center}
\textbf{\Large {Модель Поттса с тремя состояниями на дереве Кэли: периодические меры Гиббса }}\\
\end{center}

\begin{center}
Ф.Х.Хайдаров, Р.М.Хакимов
\end{center}

Теория мер Гиббса-это новая теория относительно теории мер. Одним из 
основных задач этой теории является существование фазого перехода 
той или иной физической системы. Если мера Гиббса не единственна, 
то существует фазовый переход (см.например \cite {6}). Поэтому изучение мер Гиббса
играет важную роль во многих областях науки.

В работе \cite{Ga8} изучена ферромагнитная модель Поттса с тремя
состояниями на дереве Кэли второго порядка. В работе \cite {GN}
обобщены результаты работы \cite{Ga8} для модели Поттса с конечным
числом состояний на дереве Кэли произвольного (конечного) порядка.

Для антиферромагнитной модели Поттса с внешним полем показано, что 
трансляционно-инвариантная мера Гиббса единственна  (см. \cite{R}). Работа
\cite{Ga13} посвящена модели Поттса со счетным числом состояний и
c ненулевым внешним полем.

В работе \cite{RK} изучены периодические меры Гиббса на дереве Кэли второго порядка и при некоторых условиях доказано, что все
периодические меры Гиббса являются трансляционно-инвариантными;
найдены условия, при которых модель Поттса с ненулевым внешним
полем имеет периодические меры Гиббса. Работа \cite{KhR1} является продолжением работы \cite{RK}. Доказана существование не менее трех периодических мер Гиббса с периодом два на дереве Кэли порядка три и четыре. В работе \cite
{KRK} дано полное описание трансляционно-инвариантных мер Гиббса
для ферромагнитной модели Поттса с $q$ состояниями и показано, что
их количество равно $2^q-1$, а в работе \cite{KR} изучена задача крайности этих мер.

В этой работе мы покажем, что при некоторых условиях на параметр
модели Поттса с тремя состояниями с нулевым внешним полем на
дереве Кэли порядка $k\geq 3$ существуют ровно две
периодические (не трансляционно-инвариантные) меры Гиббса.\

Дерево Кэли $\Im^k$ порядка $ k\geq 1 $ - бесконечное дерево, т.е.
граф без циклов, из каждой вершины которого выходит ровно $k+1$
ребро. Пусть $\Im^k=(V,L,i)$, где $V-$есть множество вершин
$\Im^k$, $L-$множество его ребер, и $i-$функция
инцидентности, сопоставляющая каждому ребру $l\in L$ его концевые
точки $x, y \in V$. Если $i (l) = \{ x, y \} $, то $x$ и $y$
называются  {\it ближайшими соседями вершины} и обозначается $l =
<x, y> $. Расстояние $d(x,y), x, y \in V$ на дереве Кэли
определяется формулой
$$
d (x, y) = \min \ \{d | \exists x=x_0, x_1,\dots, x _ {d-1},
x_d=y\in V \ \ \mbox {такой, что} \ \ <x_0, x_1>,\dots, <x _
{d-1}, x_d>\}.$$

Для фиксированного $x^0\in V$ обозначим $ W_n = \ \{x\in V\ \ | \ \
d (x, x^0) =n \}, $
$$ V_n = \ \{x\in V\ \ | \ \ d (x, x^0) \leq n \},\ \
L_n = \ \{l = <x, y> \in L \ \ | \ \ x, y \in V_n \}. \eqno (1) $$

Известно, что существует взаимнооднозначное соответствие между
множеством $V$ вершин дерева Кэли порядка $k\geq 1 $ и группой $G
_{k},$ являющейся свободным произведением $k+1$ циклических групп
второго порядка с образующими $a_1, a_2,\dots, a_{k+1} $,
соответственно.\

Мы рассмотрим модель, где спиновые переменные принимают значения
из множества $\Phi = \ \{1, 2,\dots, q \},$ $ q\geq 2 $ и
расположены на вершинах дерева. Тогда \emph{ конфигурация} $\s$ на
$V$ определяется как функция $x\in V\to\s (x) \in\Phi$; множество
всех конфигураций совпадает с $\Omega =\Phi ^ {V} $.

Гамильтониан модели Поттса определяется как
$$H(\sigma)=-J\sum_{\langle x,y\rangle\in L}
\delta_{\sigma(x)\sigma(y)},\eqno(2)$$ где $J\in R$, $\langle x,y\rangle-$ ближайшие соседи и
$\delta_{ij}-$ символ Кронекера:
$$\delta_{ij}=\left\{\begin{array}{ll}
0, \ \ \mbox{если} \ \ i\ne j\\[2mm]
1, \ \ \mbox{если} \ \ i= j.
\end{array}\right.
$$
Определим конечномерное распределение вероятностной меры $\mu$ в
обьеме $V_n$ как $$\mu_n(\sigma_n)=Z_n^{-1}\exp\left\{-\beta
H_n(\sigma_n)+\sum_{x\in W_n}h_{\sigma(x),x}\right\},\eqno(3)$$
где $\beta=1/T$, $T>0$--температура,  $Z_n^{-1}$ нормирующий
множитель и $\{h_x=(h_{1,x},\dots, h_{q,x})\in R^q, x\in V\}$
совокупность векторов и
$$H_n(\sigma_n)=-J\sum_{\langle x,y\rangle\in L_n}
\delta_{\sigma(x)\sigma(y)}.$$

Говорят, что вероятностное распределение (3) согласованное, если
для всех $n\geq 1$ и $\sigma_{n-1}\in \Phi^{V_{n-1}}$:
$$\sum_{\omega_n\in \Phi^{W_n}}\mu_n(\sigma_{n-1}\vee
\omega_n)=\mu_{n-1}(\sigma_{n-1}).\eqno(4)$$

Здесь $\sigma_{n-1}\vee \omega_n$  есть объединение конфигураций.
В этом случае, существует единственная мера $\mu$ на $\Phi^V$
такая, что для всех $n$ и $\sigma_n\in \Phi^{V_n}$
$$\mu(\{\sigma|_{V_n}=\sigma_n\})=\mu_n(\sigma_n).$$
Такая мера называется расщепленной гиббсовской мерой,
соответсвующей гамильтониану (2) и векторзначной функции $h_x,
x\in V$.\

Следующее утверждение описывает условие на $h_x$, обеспечивающее
согласованность $\mu_n(\sigma_n)$.

\textbf{Теорема 1}.\cite{R} \textit{Вероятностное распределение
$\mu_n(\sigma_n)$, $n=1,2,\ldots$ в (3) является согласованной
тогда и только тогда}, \textit{когда для любого} $x\in V$
\textit{имеет место следующее
$$h_x=\sum_{y\in
S(x)}F(h_y,\theta),\eqno(5)$$ где $F: h=(h_1,
\dots,h_{q-1})\in R^{q-1}\to
F(h,\theta)=(F_1,\dots,F_{q-1})\in R^{q-1}$ определяется
как:
$$F_i=\ln\left({(\theta-1)e^{h_i}+\sum_{j=1}^{q-1}e^{h_j}+1\over
\theta+ \sum_{j=1}^{q-1}e^{h_j}}\right),$$ и
$\theta=\exp(J\beta)$, $S(x)-$ множество прямых потомков точки
$x$.}\\

Пусть $\widehat{G}_k-$ подгруппа группы $G_k$.

\textbf{Определение 1}. Совокупность векторов $h=\{h_x,\, x\in
G_k\}$ называется $ \widehat{G}_k$-периодической, если
$h_{yx}=h_x$ для $\forall x\in G_k, y\in\widehat{G}_k.$

$G_k-$ периодические совокупности называются
трансляционно-инвариантными.

\textbf{Определение 2}. Мера $\mu$ называется
$\widehat{G}_k$-периодической, если она соответствует
$\widehat{G}_k$-периодической совокупности векторов $h$.

Следуюшая теорема характеризует периодические меры Гиббса.

\textbf{Теорема 2.}\cite{RK} \textit{Пусть $K-$ нормальный
делитель конечного индекса в $G_k.$ Тогда для модели Поттса все
$K-$ периодические меры Гиббса являются либо $G_k^{(2)}-$
периодическими, либо трансляционно-инвариантными.}\

Рассмотрим случай $q=3$, т.е.
$\sigma:V\rightarrow\Phi= \{1,2,3\}$. В силу Теоремы 2 имеются
только $G^{(2)}_k$-периодические меры Гиббса, которые
соответствуют совокупности векторов $h=\{h_x\in R^{q-1}: \, x\in
G_k\}$ вида
$$h_x=\left\{%
\begin{array}{ll}
    h^1, \ \ \ $ если $ |x|-\mbox{четно} $,$ \\
    h^2, \ \ \ $ если $ |x|-\mbox{нечетно} $.$ \\
\end{array}%
\right. $$
 Здесь $h^1=(h_1^1,h_2^1),$ $h^2=(h_1^2,h_2^2).$
Тогда в силу (5) имеем:
$$
\left\{%
\begin{array}{ll}
    h_{1}^1=k\ln\left({\theta\exp(h_1^2) + \exp(h_2^2)+1\over \exp(h_1^2) + \exp(h_2^2)+\theta}\right)\\[3 mm]
    h_{2}^1=k\ln\left({\exp(h_1^2) +\theta \exp(h_2^2)+1\over \exp(h_1^2) + \exp(h_2^2)+\theta}\right) \\[3 mm]
    h_{1}^2=k\ln\left({\theta\exp(h_1^1) + \exp(h_2^1)+1\over \exp(h_1^1) + \exp(h_2^1)+\theta}\right) \\[3 mm]
    h_{2}^2=k\ln\left({\exp(h_1^1) + \theta\exp(h_2^1)+1\over \exp(h_1^1) + \exp(h_2^1)+\theta}\right). \\
\end{array}%
\right.$$\

Введем следующие обозначения: $\exp(h_1^1)=z_1,\ \exp(h_2^1)=z_2,
\ \exp(h_1^2)=z_3, \ \exp(h_2^2)=z_4.$ Тогда последнюю систему
уравнений можно переписать:
$$
\left\{%
\begin{array}{ll}
    z_{1}=\left({\theta z_3 + z_4+1\over z_3 + z_4+\theta}\right)^k \\[3 mm]
    z_{2}=\left({\theta z_4 + z_3+1\over z_3 + z_4+\theta}\right)^k \\[3 mm]
    z_{3}=\left({\theta z_1 + z_2+1\over z_1 + z_2+\theta}\right)^k \\[3 mm]
    z_{4}=\left({\theta z_2 + z_1+1\over z_1 + z_2+\theta}\right)^k. \\
\end{array}%
\right.\eqno(6)$$\

Для решений системы уравнений (6) верно следующее

\textbf{Утверждение.} \textit{При любом $k$ и $J<0(\theta<1)$ справедливы
следующие утверждения:}

1. $z_1<z_2(z_1>z_2)$ \textit{тогда и только тогда, когда}
$z_3>z_4(z_3<z_4).$

2. $z_1\leq1(z_1\geq1)$ \textit{тогда и только тогда, когда}
$z_3\geq1(z_3\leq1).$

3. $z_2\leq1(z_2\geq1)$ \textit{тогда и только тогда, когда}
$z_4\geq1(z_4\leq1).$

\textbf{Доказательство.} 1. В системе уравнений (6) вычтем из
первого уравнения второе:
$$z_1-z_2=\frac{(\theta-1)(z_3-z_4)}{z_3+z_4+\theta}C,$$
где
$$C={\left(\theta z_3 + z_4+1\over z_3 + z_4+\theta\right)^{k-1}}+\ldots+{\left(\theta z_4 + z_3+1\over z_3 + z_4+\theta\right)^{k-1}}.$$
Отсюда т.к. $C>0, z_3+z_4+\theta>0,$ то при  $0< \theta<1$ получим требуемое.

2. Пусть $z_1\leq1$(случай $z_1\geq 1$ доказывается аналогично).
Тогда из первого уравнения системы (6) имеем
$z_3(\theta-1)\leq\theta-1.$ Так как $0<\theta <1,$ то из последнего
получим, что $z_3\geq 1.$

Пусть теперь $z_3\geq 1.$ Тогда из третьего уравнения системы (6)
будем иметь $z_1(\theta-1)\geq \theta-1.$
Отсюда при $0<\theta <1$ получим, что $z_1\leq 1.$

3. Доказывается аналогично предыдущему. Утверждение доказано.

Известно, что $I=\{z\in R^4: z_1=z_2, \ z_3=z_4\}$-инвариантное
множество отображения (6)(см.\cite{RK}). Покажем, что система
уравнений (6) при $J<0$ $(0<\theta<1)$ и произвольного $k\geq 3$
имеет неединственное решение $(z_1,z_2,z_3,z_4)$ с $z_1=z_2=x$,
$z_3=z_4=y$. При этих условиях из (6) имеем

$$
\left\{%
\begin{array}{ll}
    x=f(y) \\
    y=f(x), \\
    \end{array}%
\right. \texttt{где}\ f(x)=\left[{(\theta +1)x+1\over 2x
+\theta}\right]^k. \eqno(7)$$\

Пусть $\theta_{cr}\equiv \theta_{cr}=\frac{k-2}{k+1}$. Тогда верна
следующая

\textbf{Теорема 3.} \textit{Для модели Поттса } \textit{при}
$k\geq 3$,\, $q=3$, \, $J<0$ \textit{при $0<\theta<\theta_{cr}$ существуют ровно три $G_k^{(2)}-$
периодические меры Гиббса, соответствующие совокупности из множества $I$. При этом одна из них является трансляционно-инвариантной, а другие две $G_k^{(2)}-$
периодическими (не трансляционно-инвариантными).}\

\textbf{Доказательство.} Легко увидеть, что система уравнений (7) имеет решение $(1,1)$, которое соответствует единственной трансляционно-инвариантной мере Гиббса $\mu_1$. Докажем, что при условиях теоремы сущуствуют только две $G_k^{(2)}-$ периодические (не трансляционно-инвариантные) меры Гиббса, отличные от меры $\mu_1$. Для этого из (7) получим уравнение $x=f(f(x))$. Так как функция $f(x)$ обратима при $x>0$, то последнее уравнение можно изучить в следующем виде  $f(x)=f^{-1}(x)=g(x)$. Рассмотрим функцию $h(x)=\ln{f(x)\over g(x)}=\ln f(x)-\ln g(x)$ и вычислим производные
$$f'(x)={k(\theta-1)(\theta+2)f(x)\over ((\theta+1)x+1)(2x+\theta)}, \ \ g'(x)={(\theta-1)(\theta+2)g(x)\over k\sqrt[k]{x^{k-1}}(2\sqrt[k]{x}-\theta-1)(1-\theta\sqrt[k]{x})}$$
и пользуясь этими
$$h'(x)={f'(x)\over f(x)}-{g'(x)\over g(x)}={(\theta-1)(\theta+2)\over k}\left({k^2\over ((\theta+1)x+1)(2x+\theta)}-{1\over \sqrt[k]{x^{k-1}}(2\sqrt[k]{x}-\theta-1)(1-\theta\sqrt[k]{x})}\right).$$
Ясно, что $h(1)=0$ и из условия $h'(1)<0$ вытекает $\theta<{k-2\over k+1}=\theta_{cr}.$ Из условия $g(x)>0$ при $0<\theta<{k-2\over k+1}<1$ имеем
$$\theta_1=\left({\theta+1\over2}\right)^k<x<{1\over \theta^k}=\theta_2.$$
Обозначив $\sqrt[k]{x}=y$, перепишем производную $h'(x)$ следующим образом
$$\alpha(y)={(\theta-1)(\theta+2)p(y)\over k\left(2(\theta+1)y^{2k}+(\theta^2+\theta+2)y^k+\theta\right)\left(2\theta y^{k+1}-(\theta^2+\theta+2)y^k+(\theta+1)y^{k-1}\right)},$$
где
$$p(y)=2(\theta+1)y^{2k}+2\theta k^2y^{k+1}-(k^2-1)(\theta^2+\theta+2)y^k+k^2(\theta+1)y^{k-1}+\theta.$$
Отсюда многочлен $p(y)$ и значит $h'(x)$ имеет не более двух положительных решения по теореме Декарта о положительных корней многочлена. Легко проверить, что $$\lim_{x\rightarrow\theta_1}h'(x)=+\infty, \ \lim_{x\rightarrow\theta_2}h'(x)=+\infty$$
при $\theta<\theta_{cr}$. Тогда из $h'(1)<0$ следует, для функции $h(x)$ существуют ровно две критические точки $\theta_1<\xi_1<1$ и $1<\xi_2<\theta_2$. Кроме того,
$$\lim_{x\rightarrow\theta_1}h(x)=-\infty, \ h(1)=0, \ \lim_{x\rightarrow\theta_2}h(x)=+\infty.$$
Значит, $h(x)$ возрастает при $\theta_1<x<\xi_1, \ x>\xi_2$ и убывает при $\xi_1<x<\xi_2$. Следовательно, уравнение  $h(x)=0$ имеет только три решения $x_0<x_1=1<x_2$, т.е. для модели Поттса при условиях теоремы существуют только три $G_k^{(2)}-$периодические меры Гиббса, одна из которых является трансляционно-инвариантной, а две другие являются $G_k^{(2)}-$
периодическими (не трансляционно-инвариантными).Теорема доказана.\\

\newpage

\begin{center}Ф.Х.Хайдаров \\ Национальный университет Узбекистана.

email: haydarov\_imc@mail.ru
\end{center}

\begin{center}Р.М.Хакимов \\ Институт математики,
29, ул. Дурмон йули, 100125, Ташкент.

email: rustam-7102@rambler.ru
\end{center}

\end{document}